\documentclass[12pt]{article}
\usepackage[utf8]{inputenc}
\usepackage{amsmath}
\usepackage[margin=1.5in]{geometry}
\usepackage{amsthm}
\usepackage{parskip}
\usepackage{xcolor}
\usepackage{float}
\definecolor{ourblue}{HTML}{0571a3}
\usepackage[colorlinks=true, allcolors=ourblue]{hyperref}
\usepackage{titling}
\usepackage{authblk}
\usepackage{caption}

\usepackage{setspace}
\usepackage{mathptmx}

\newenvironment{journalabstract}{\section*{Abstract}}
{}

\DeclareCaptionLabelFormat{mid}{#1 #2 $\mathbf{\vert}$~}

\newcommand{\SupplementNameFull}{Supplementary Information}

\newcommand{\papertitle}{Structured foraging of soil predators\\ unveils functional responses to bacterial defenses}
\title{\papertitle}

\author[1*]{Fernando W.~Rossine}
\author[2*]{Gabriel Vercelli}
\author[1]{Corina E.~Tarnita}
\author[2,3]{Thomas Gregor}
\affil[1]{Department of Ecology \& Evolutionary Biology, Princeton University, Guyot Hall, Princeton, NJ 08544, USA}
\affil[2]{Joseph Henry Laboratory of Physics  
  \& Lewis-Sigler Institute for Integrative Genomics,
  Princeton University, Princeton, NJ 08544, USA}
\affil[3]{ Department of Stem Cell and Developmental Biology, CNRS UMR3738,
 Institut Pasteur, 25 rue du docteur Roux, 75015 Paris, France}
\affil[*]{These authors contributed equally}

\usepackage[backend=biber, style=nature]{biblatex}
\usepackage{graphicx}
\usepackage{amsfonts}

\usepackage{physics}
\usepackage{appendix}

\usepackage{macros/optimal-theorem-environments}
\usepackage{macros/optimal-macros} 
\newcommand{\autocaption}[2]{\caption[#1]{\textbf{#1} #2}}

\begin{document}
\maketitle

\bigskip
\begin{journalabstract}
 Predators and their foraging strategies often determine ecosystem structure and function. Yet, the role of protozoan predators in microbial soil ecosystems remains elusive despite the importance of these ecosystems to global biogeochemical cycles. In particular, amoebae --- the most abundant soil protozoan predators of bacteria --- remineralize soil nutrients and shape the bacterial community. However, their foraging strategies and their role as microbial ecosystem engineers remain unknown. 
 Here we present a multi-scale approach, connecting microscopic single-cell analysis and macroscopic whole ecosystem dynamics, to expose a phylogenetically widespread foraging strategy, in which an amoeba population spontaneously partitions between cells with fast, polarized movement and cells with slow, unpolarized movement. Such differentiated motion gives rise to efficient colony expansion and consumption of the bacterial substrate. From these insights we construct a theoretical model that predicts how disturbances to amoeba growth rate and movement disrupt their predation efficiency. These disturbances correspond to distinct classes of bacterial defenses, which allows us to experimentally validate our predictions. All considered, our characterization of amoeba foraging identifies amoeba mobility, and not amoeba growth, as the core determinant of predation efficiency and a key target for bacterial defense systems.
\end{journalabstract}

\newpage

\section*{Introduction}

Throughout biomes and across spatial scales --- from the vast African savannas to the minute rhizosphere microcosms --- foraging strategies are key to how predators shape species diversity and ecosystem function\autocite{estes2011trophic}. Studies of macroscopic ecosystems have revealed that different aspects of predator foraging strategies --- e.g., exploration behavior, dietary selectivity --- determine prey encounter rates and the effectiveness of predation deterrents such as plant secondary metabolites\autocite{iason2006behavioral,foley2005plant} or porcupine quills\autocite{katzner2015most}. Yet, microbial foraging strategies --- in particular those of soil bacterivore protozoans --- remain understudied due to unique challenges: the opacity of soil impedes \textit{in loco} behavioral observation\autocite{brethauer2020impacts}, and metagenomic methods are hindered by the substantial amplification biases in protozoan sequences\autocite{geisen2018soil} and the low abundance of protozoan predators relative to their bacterial prey\autocite{crowther2019global}. All told, we do not know what strategies soil protozoan predators employ to explore their microscopic landscape, nor do we understand the consequences of such foraging strategies to bacterial population dynamics or to the effectiveness of bacterial defenses.

Thus, characterizing foraging strategies is a crucial missing link to a predictive mechanistic understanding of the well-documented ecological importance of soil protozoan predators, of which amoebae are the most widespread and abundant\autocite{bates2013global,adl2005dynamics,clarholm1981protozoan}. The amoebae are generalist predators that mold soil communities by consuming and limiting bacterial populations\autocite{clarholm1981protozoan,rosenberg2009soil,kreuzer2006grazing} all the while altering bacterial taxonomic diversity\autocite{rosenberg2009soil,kreuzer2006grazing}. Ultimately, amoebae consume and remineralize a significant portion of the total soil microbial productivity\autocite{de1994modelling,elliott1977soil}, leading to increased nutrient cycling\autocite{bonkowski2004protozoa,clarholm1985interactions}. But the consequences of microbial predation go far beyond the soil itself, ranging from changes to biogeochemical cycles to the emergence of pathogens: predator-driven increases of bio-available nitrogen fertilize plants\autocite{bonkowski2004protozoa,kreuzer2006grazing,clarholm1985interactions}, remineralized carbon returns to the atmosphere\autocite{bonkowski2004protozoa}, and the interactions between amoebae and fungi are implicated in the evolution of fungal pathogenicity\autocite{albuquerque2019hidden}.

Amoebae have a molecular machinery that enables prey search and handling behaviors required for complex foraging strategies. Amoeba cells can perceive and integrate various cues — mechanical forces\autocite{riviere2007signaling}, temperature\autocite{hong1983thermotaxis}, and attractant/repellent chemical gradients\autocite{bonner1969acrasin,phillips2012secreted} — leading to oriented cell movement, often towards prey. Once amoebae encounter bacteria they deploy a lineup of lectins that bind to specific bacteria surface carbohydrates, leading to selective prey handling and consumption\autocite{farinholt2019microbiome}. When scaled up to thousands of amoebae, such multi-factorial behaviors may lead to complex emergent features of the predator-prey interaction. Indeed, studies have suggested that amoebae at higher densities consume bacteria more efficiently, pointing towards benefits of collective prey-handling\autocite{rubin2019cooperative}. The locomotive behavior of a collection of cells may also lead to emergent ecological properties. For instance, experimental and theoretical work on bacteria have shown that the interplay between nutrient consumption and cell movement orientation creates chemical gradients that determine how the bacterial population expands into uncolonized regions\autocite{cremer2019chemotaxis,narla2021traveling}. Analogously, other studies have suggested that the secretion of chemorepellents might play a role in the expansion of soil amoebae colonies\autocite{phillips2012secreted,phillips2010roco}. Overall, describing how the behavior of individual amoeba cells results in emergent foraging strategies is fundamental to understanding the efficiency with which amoebae deplete their bacterial prey.

In this study, we identify the foraging strategies that soil amoebae adopt when invading a spatially structured bacterial matrix and show that such foraging strategies determine both the rates of bacteria consumption as well as the effectiveness of different functional classes of bacterial anti-predator defenses. We employ a cross-scale approach, characterizing both the microscopic movement patterns of single amoebae within the bacterial matrix, and the emergent macroscopic spatial patterns of predation. We first investigate the foraging strategy of \textit{Dictyostelium discoideum}, a well studied bacterivore soil amoeba species with a complex life cycle comprised of a unicellular foraging stage and a multi-cellular dispersal stage\autocite{bonner2015cellular}. To assess the generality of the observed patterns, we compare \textit{D. discoideum}’s foraging strategy with that of other soil amoebae species with distinct life cycles. In all investigated species we find cell behavior differentiation and the occurrence of coordinated cell movement. We build a mathematical model that integrates our microscopic and macroscopic observations to make predictions about the macroscopic consequences of disturbing individual cell behaviors. We test these predictions using chemical deterrents that mimic two functional classes of bacterial defenses --- predator movement deterrents and predator growth rate deterrents. Altogether, we present an experimental and theoretical framework that quantitatively bridges individual amoeba behaviors and self-organized growth, predator-prey dynamics, and microbial defenses.

\bigskip
\section*{Results}
\subsection*{Macroscopic and microscopic features of  bacterial matrix invasion}

\begin{figure}[t]
    \centering
\makebox[\textwidth][c]{    \includegraphics[width=6.5in]{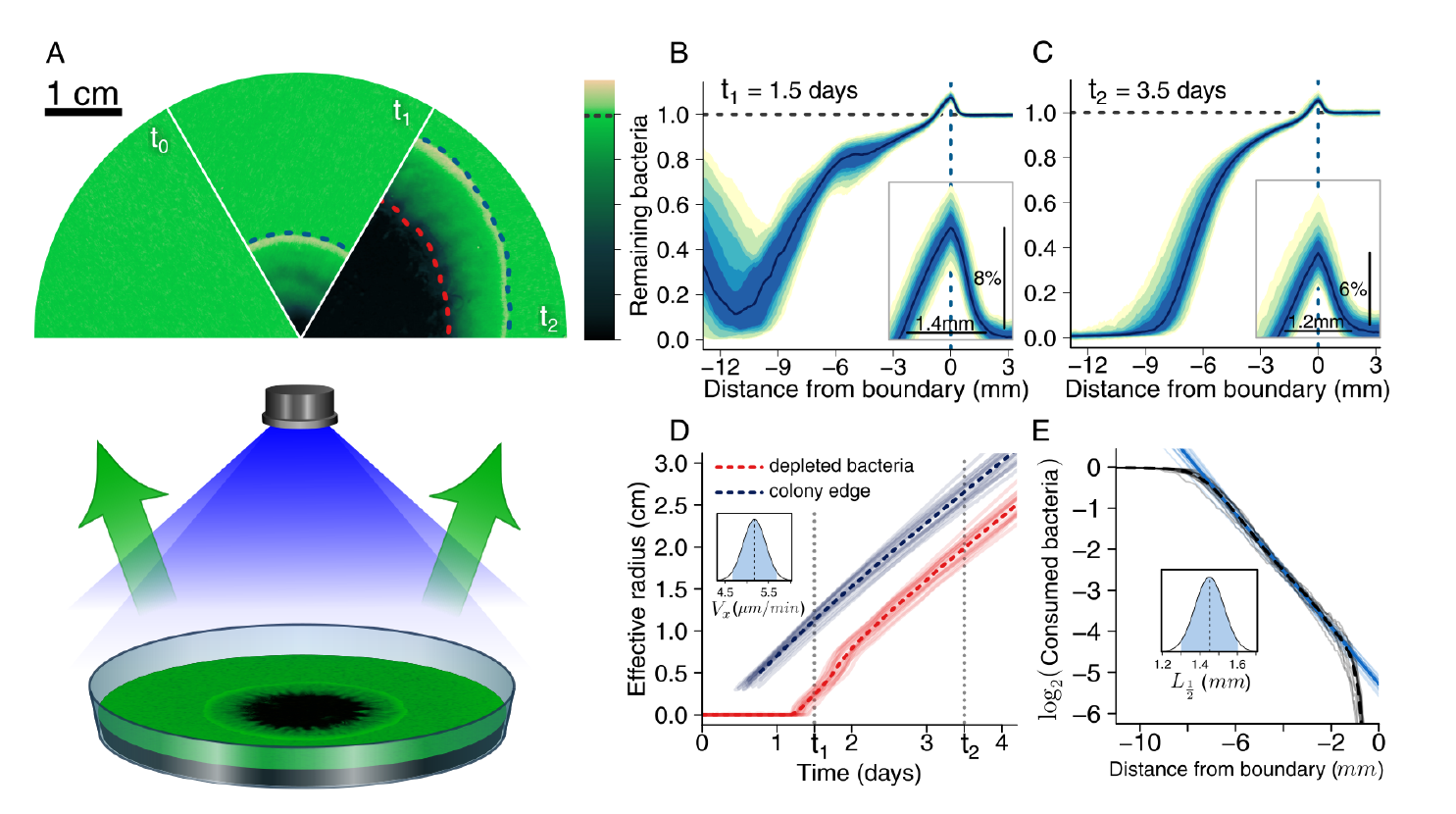}}
\autocaption{Macroscopic features of \dicty  invasion:}{\textbf{A}, Schematic shows a \dicty colony growing in a bacterial lawn illuminated by a blue LED projector with EGFP-labelled \ecoli emitting green light. Sectors show the same Petri dish at $t_0=0$, $t_1=1.5$, and $t_2=3.5$ days after amoeba inoculation. Dashed blue line denotes amoeba colony boundary and dashed red line denotes area of bacterial exhaustion. False color denotes proportion of remaining bacteria. \textbf{B-E}, Show data collected for 16 Petri dishes. \textbf{B-C}, Aggregate profiles of remaining bacteria as a function of distance from the colony boundary for times $t_1$ (\textbf{B}) and $t_2$ (\textbf{C}), each calculated from 40 profiles taken from each of the 16 dishes. Insets show detail of bacterial accumulation along colony boundary. \textbf{D}, Time courses for the effective radii for both the colony and the depleted bacteria zone. Inset shows estimated values of $\Vx$. \textbf{E}, Logarithm of consumed bacteria proportion as a function of distance from colony boundary. Inset shows estimated values of $\Lhalf$. For (\textbf{D,E}) transparent lines show time courses for individual dishes and dashed lines show mean time course. }
\label{fig:macro}
\end{figure}

To elucidate the dynamics of how \dicty invades the bacterial matrix we first set out to characterize the macroscopic features of such an invasion. Notably, to ensure that our observations reflect the natural behavior of \dicty cells, we use natural isolates, rather than the more commonly used lab strains, known for their anomalous colony morphologies\autocite{fey2013one}. We prepared Petri dishes with homogeneous lawns of eGFP-expressing \ecoli cells grown for two days until reaching the stationary phase. At the center of these lawns we inoculated about 100 cells of \dicty. These amoeba cells consumed bacteria, divided, and moved, thereby establishing outwardly expanding colonies (Fig. \ref{fig:macro}-\textbf{A}).

Early on during the bacterial matrix invasion, a highly fluorescent ring develops around the initial \dicty inoculation zone due to an accumulation of bacteria. This luminous ring demarcates the boundary of the feeding front beyond which amoeba cells are not present. Thus the luminous ring is used to track the expansion of the nascent amoeba colony (Fig. \ref{fig:macro}-\textbf{A}). We find that \dicty has colony expansion rates of $V_x=5.2\pm 0.3 \,{\rm \mu m/min}$ on average. Notably, colonies attain such expansion rates from the moment that the luminous ring becomes visible and before any substantial bacterial consumption occurs (Fig. \ref{fig:macro}-\textbf{D}). This is surprising because classical models for expanding populations of isotropically moving cells predict that colony expansion rates should accelerate until the amoeba feeding front stabilizes, which can only occur after bacteria are depleted at the center of the dish\autocite{murray2007mathematical}. This discrepancy between classical models and our observations suggests that anisotropic cell movement might be an important factor in \dicty colony expansion.

To further characterize the feeding front, we traced radial fluorescence intensity profiles, normal to the colony boundary (Fig. \ref{fig:measuring}-\textbf{A,B}). Because fluorescence is lost when bacteria are consumed by amoebae (Fig. \ref{fig:measuring}-\textbf{C,D}), these profile measurements quantify the proportion of consumed bacteria as a function of the distance from the colony boundary. After a transient period lasting up to 2 days, the bacteria consumption profile adopts a constant shape consisting of three distinct zones: the \dicty colony boundary, which is marked by a local accumulation of bacteria; a zone of exponential consumption of bacteria; and a zone exhausted of bacteria (Fig. \ref{fig:macro}-\textbf{B,C}). To characterize the exponential growth of the amoeba colony, we introduce the halving length (\Lhalf ) of the feeding front. It measures the characteristic distance across the feeding front such that the proportion of consumed bacteria falls by half. For \dicty we measured halving lengths of $\Lhalf=1.45\pm 0.08\,{\rm mm}$ (Fig. \ref{fig:macro}-\textbf{E}).

$\Vx$ and $\Lhalf$ provide a temporal and a spatial descriptor, respectively, for the macroscopic characterization of amoeba feeding fronts: $\Vx$ reveals how the feeding front advances in time and $\Lhalf$ reveals how the feeding front is shaped (Fig. \ref{fig:measuring}-\textbf{E,F}). Importantly, $\Vx$ and $\Lhalf$ are not just descriptors of the feeding front, but are also quantities that underlie ecological aspects of the predator-prey interaction: as the bacteria consumption profile stabilizes, $\Vx$ sets the global bacteria consumption rate (i.e. ), while $\Lhalf$ sets the area over which amoeba cells compete for resources by defining the spatial extent of the feeding front.

\begin{figure}[ht!]
    \centering
\makebox[\textwidth][c]{    \includegraphics[width=6.5in]{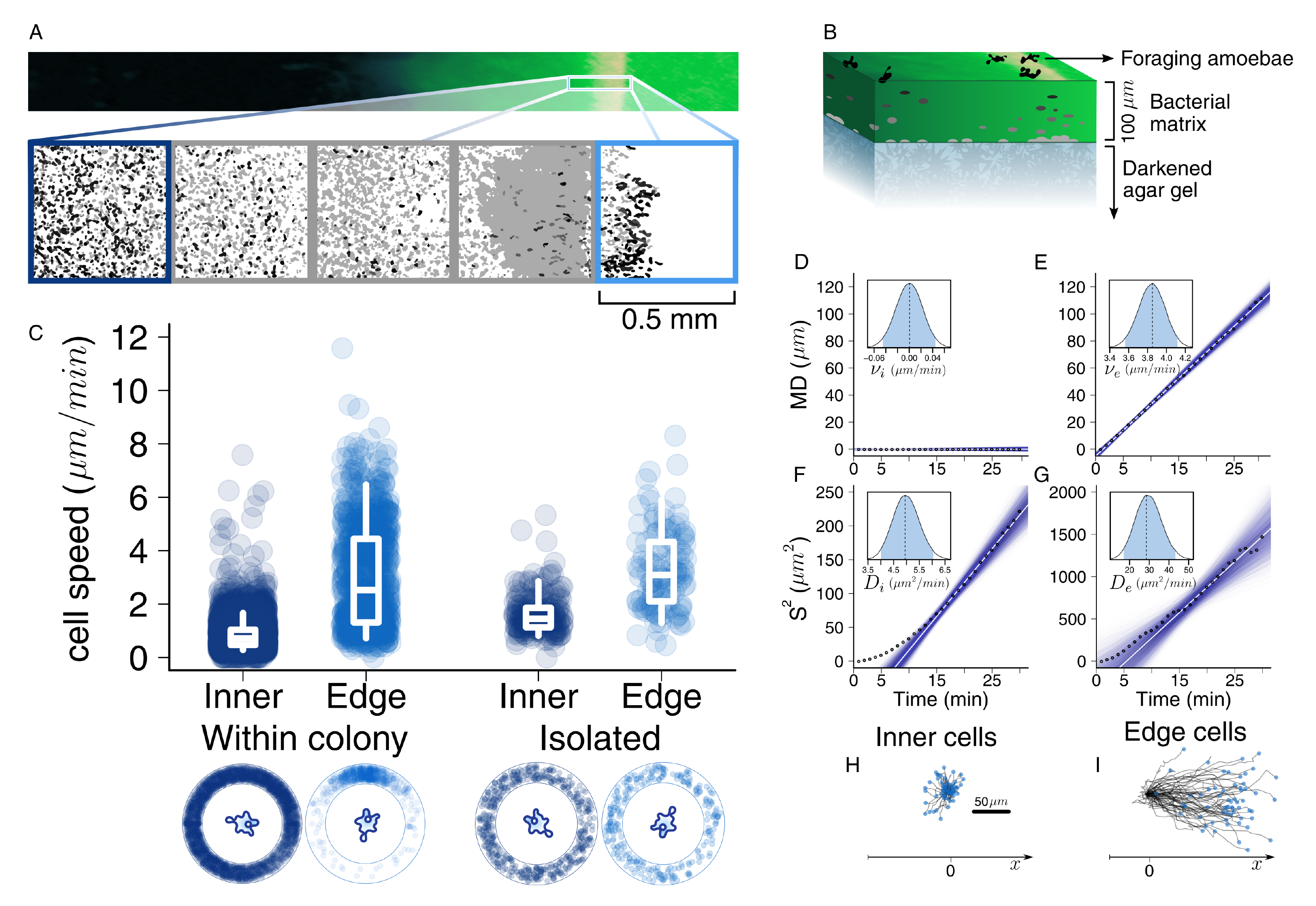}}
\autocaption{Microscopic features of \dicty  invasion:}{\textbf{A}, Region of the feeding front imaged with a confocal microscope. Expanded region shows tresholded confocal images showing \dicty cells at different depths of the bacterial matrix, with cells shown in black at the surface of the bacterial matrix and cells shown in light gray at the bottom. Each box corresponds to one field of view along the feeding front, ranging from the inner cells (dark blue) to the edge cells (light blue). \textbf{B}, Schematic of the agar slab with the feeding front showing amoeba cells at different depths. \textbf{C}, Median speed (box plots) and direction (ring plots) for individual amoeba trajectories, for inner and edge cells, either embedded in the native feeding front or isolated from the colony. Whiskers of the box plots denote the $5^{th}$ and $95^{th}$ percentiles. \textbf{D-G}, Mean amoeba displacement (\textbf{MD}) and position variance (S2???) are shown as a function of time for inner and edge cells. Black points denote data, blue lines are bootstrapped linear regressions (after a relaxation time) and white lines are the median regressions. Insets show bootstrapped estimates of advection and diffusion coefficients. \textbf{H-I}, Sample trajectories of inner (\textbf{H}) and edge (\textbf{I}) cells.}
\label{fig:micro}
\end{figure}

We next set out to derive the observed macroscopic features of the bacterial matrix invasion from the underlying microscopic organization and dynamics of amoeba cells at the feeding front. To this end, we cut out agar slabs around the feeding front, and placed them under a confocal microscope. Infusion of the bacterial lawn with fluorescence (i.e. fluorescein), allows us to track amoeba cells as dark regions within the fluorescent background (Fig. \ref{fig:micro}-\textbf{A,B}). While amoeba cells were present at all depths of the approximately $100\,{\rm \mu m}$ thick bacterial matrix, most amoeba cells are located either at its surface or at the interface between the agar gel and the bacterial matrix. At the very boundary of the feeding front, marked by the macroscopically observed luminous ring, amoeba cells accumulate --- both at the surface of the bacterial matrix but also at the interface with the agar gel --- forming a sharp band (Fig. \ref{fig:micro}-\textbf{A}). We consider two amoeba cell populations in particular: cells around the boundary of the feeding front (referred to as edge cells) and cells $2.5\,{\rm mm}$ behind the boundary (referred to as inner cells). Tracking individual cells of these two populations uncovers two strikingly distinct behavioral patterns: inner cells moved in no preferential direction, whereas edge cells displayed markedly polarized trajectories, moving preferentially away from the colony center (Fig \ref{fig:micro} \textbf{H,I}). Moreover, we find that with an average speed of $\bar{v}_e=3.00\pm 0.06\,{\rm \mu m/min}$ edge cells move over three times faster than inner cells with $\bar{v}_i=0.81\pm 0.01\,{\rm \mu m/min}$ (Fig. \ref{fig:micro}-\textbf{C}).

Polarized movement of individual amoeba cells suggests that they respond to some environmental cue (e.g. a bacterially produced chemoattractant, a chemorepellent produced by amoeba cells, or rheological changes to the bacterial matrix caused directly or indirectly by the amoeba cells). Such cues could in fact be responsible for the observed difference between edge- and inner-cell speeds: in many chemotaxis systems both, cell polarization and higher cell speed, arise via a common response to chemical gradients\autocite{armitage1997bacterial,keller1971model}. However, it is unclear whether differences in polarization and speed are indeed linked. To test in our system whether cell speed differences persist in the absence of environmental cues, we collect edge cells and inner cells separately, washed them, and placed them in buffer-filled glass bottom dishes. As expected, as cells settled and resumed their pseudopodial motion in this new environment that is devoid of directional cues, no polarized cell movement could be detected. However, edge cells remained faster than inner cells, indicating that some of the behavioral differentiation between inner and edge cells can, at least transiently, persist in the absence of environmental cues (Fig. \ref{fig:micro}-\textbf{C}).

The presence of individuals with high and low mobility in the same population is a common pattern in ecology, showing up in species as disparate as sea slugs\autocite{krug2009not}, insects\autocite{roff1991wing}, bacteria\autocite{armitage1997bacterial}, and even in the multicellular stage of the slime mold life cycle\autocite{rossine2020eco}. This form of behavioral differentiation often reflects the ubiquitous tradeoff between exploiting local resources and exploring the landscape\autocite{mehlhorn2015unpacking}. Given the generality of this tradeoff, it is natural to ask whether other species of soil amoebae would adopt similar bacterial matrix invasion strategies. To investigate this question, we selected another dictyostelid, \textit{Polysphondylium violaceum}, and a more distantly related amoeba species, \textit{Acanthamoeba castellanii}, belonging to the Lobosa group\autocite{cavalier2015multigene}. Despite being soil bacterivores, all three of our species of interest are highly divergent and have very different life cycles and cell characteristics. Surprisingly, we find that both \textit{P. violaceum} and \textit{A. castellanoo} display the same general macroscopic and microscopic invasion features that we uncovered in \textit{D. discoideum} (Fig. \ref{fig:acan_poly}). We observe in all species the formation of a broad feeding front demarcated by a bacteria-dense ring, a band of amoeba cells at the colony boundary, and behavioral differentiation between slow-moving, unpolarized inner cells and fast-moving, polarized edge cells.

\subsection*{A microscopically-informed model for colony expansion}

With both macroscopic and microscopic characterizations of the amoeba colony expansion and bacterial matrix invasion at hand, we are now in a position to ask whether our microscopic description of individual amoeba cell dynamics can explain the macroscopic features of the invasion. In what follows we will use our microscopic data to build a model that quantitatively predicts the experimentally-measured macroscopic outcomes of the bacterial matrix invasion and of the amoebal colony expansion.

We model radial amoeba density $\rho(x,t)$ as a continuous quantity, taking into account cell division and movement, where $x$ is the position measured relative to the amoeba colony boundary $b$. We assume amoebae divide at a constant rate $r$ (given as a function of the mean doubling time \textsf{DT}) until bacterial prey is locally exhausted, whereupon cell division halts. We represent amoeba cell movement using a Fokker-Planck operator that consists of an advection term to capture the polarized aspect of cell movement, and a diffusive term to capture the dispersive aspect of cell movement (see SI). A crucial difference between our model and classical taxis models\autocite{keller1971model,cremer2019chemotaxis,narla2021traveling} our omission of a mechanistic description of cell movement polarization. Instead, as amoebae at different positions of the feeding front have different characteristic movement behaviors (see above), we allow diffusion and advection coefficients to be a function of the distance between the amoebae and the colony boundary. We use our edge cell tracks to extract advection ($\ve$) and diffusion ($\De$) values for cells around and beyond the colony boundary, and our inner cell tracks to extract advection ($\vi$) and diffusion ($\Di$) values for cells in the exponential consumption zone of the feeding front (Fig. \ref{fig:micro}-\textbf{D-I}). Notably, given that inner cell movement is unpolarized, the estimated value for inner cell advection is close to zero (Fig. \ref{fig:micro}-\textbf{D}). To obtain advection and diffusion values for the transitional zone (of length \textit{l}) between the colony boundary and the exponential zone, we interpolate the behavior of edge and inner cells. Following our observation that inner and edge cells retain their differential behaviors even when removed from the feeding front, we introduce a correlation time $\tHat$ for which cells retain their movement pattern (see SI).

Assuming the existence of a stable invading wave solution, we find analytical relations between the microscopic parameters of the model (i.e. $\ve$, $\vi$, $\De$, $\Di$, and \textsf{DT}) and the emergent macroscopic feeding front features (i.e. $\Vx$ and $\Lhalf$):
\begin{equation}
\begin{aligned}
\label{eqn:relationsVx}
\Vx = \ve + 2\cdot \sqrt{\frac{\De \cdot \log (2)}{\textsf{DT}}}
\end{aligned}
\end{equation}
\begin{equation}
\begin{aligned}
\label{eqn:relationsLhalf}
\Lhalf = \frac{V_x\cdot \textsf{DT}}{2} \cdot \left( 1+\sqrt{1-\frac{D_i\cdot 4\log(2)}{V_x^2\cdot \textsf{DT}}}\right).
\end{aligned}
\end{equation}
Crucially, these analytical relations are independent of the details and parameters (\textit{l}, $\tHat$) of the behavioral interpolation across the transitional zone. We can now employ these relations for an objective criterion to evaluate the performance of our model: we simply require that equations \ref{eqn:relationsVx} and \ref{eqn:relationsLhalf} hold for some combination of parameters within the ranges of our measurements or literature-obtained values.

\begin{figure}[ht]
    \centering
\makebox[\textwidth][c]{    \includegraphics[width=6.5in]{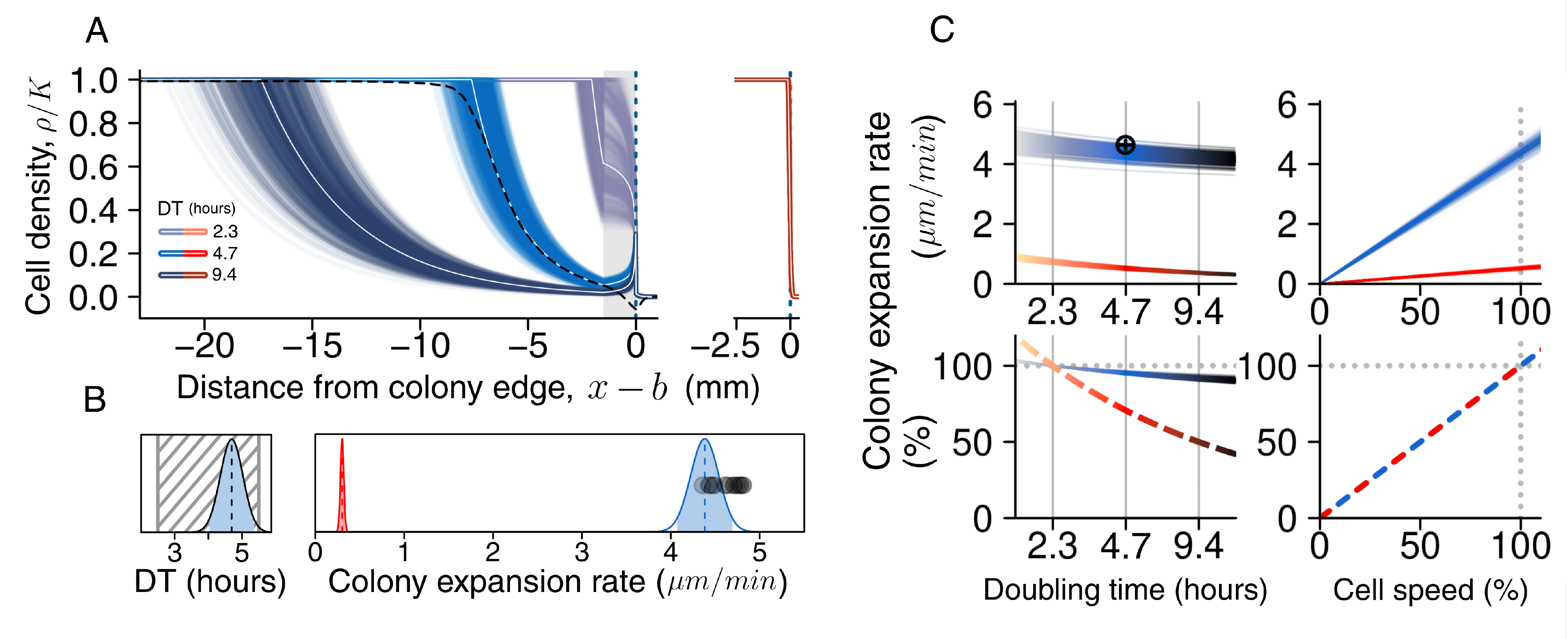}}
\autocaption{Modeling colony expansion:}{\textbf{A-C}, Results are shown for the complete model with polarized edge cells (in blue) and for the model with unpolarized edge cells (in red). \textbf{A}, Numerical realizations of the invading amoeba colony profile produced by our model for the estimated ranges of $\ve$, $\De$, $\Di$, and a fit $K$. White lines show profiles for median estimated parameters. Dashed black line shows data for bacterial consumption. \textbf{B}, Black lined bell curve shows distribution of \textsf{DT} estimated from $\Di$, $\Vx$, and $\Lhalf$, and stripped area shows a literature range for \textsf{DT}. In parallel $\Vx$ is estimated from $\ve$, $\De$, and \textsf{DT}. Black points show measured values for $\Vx$. \textbf{C}, Model predicted changes to $\Vx$ both in absolute terms (first row) or in percentage (second row) as we change cell doubling time (first column), or cell speed (second column).}
\label{fig:theo}
\end{figure}

Using this criterion, we ask whether a single type of cell movement behavior is able to account for the observed macroscopic features. We find that if inner and edge cells all behave in the same way, either all as inner cells or all as edge cells --- which we implemented by setting $\De =\Di$ and $\ve =\vi$ --- then the model cannot reconcile our macroscopic and microscopic measurements (Fig. \ref{fig:agreement} \textbf{A,B}). However, when we integrate both of the identified cell movement modes into our model, we find good agreement between macroscopic and microscopic measurements (Fig. \ref{fig:agreement} \textbf{C}). This implies that the macroscopic patterns of cell expansion cannot be attributed to any particular cell movement mode. Rather, these patterns emerge from the interplay between the distinct edge and inner cell behaviors.

Our model thus recapitulates a series of experimental observations. We find that a narrow region of amoebae cells with polarized movement produce a broad feeding front (Fig. \ref{fig:theo}-\textbf{A}), and our microscopic data accurately predicts the expansion rate of this feeding front (Fig. \ref{fig:theo}-\textbf{B}). Moreover, by taking our cell movement measurements and macroscopic colony measurements we employ equations \ref{eqn:relationsVx} and \ref{eqn:relationsLhalf} to estimate cell doubling times, which fall well within the range found in the literature (Fig. \ref{fig:theo}-\textbf{B}). Surprisingly, despite the simplicity of our assumptions about cell movement and cell division, the numerical solutions of our model produce an accumulation of amoeba density at the boundary of the colony (Fig. \ref{fig:theo}-\textbf{A}), which correspond to the sharp band of amoebae that we experimentally identified.

To further investigate which factors of cell behavior lead to the formation of the sharp amoeba band, we identified analytical conditions for the existence of such a band in our model. The emergence of the invasion band can be understood as a tug of war between edge cell movement and doubling times. Edge cells with faster, more polarized motion --- and therefore a higher $\ve$ --- favor the formation of the band by separating edge cells from the lagging exponential zone, whereas faster doubling times counteract the separation of the band by increasing cell density along the transitional zone (Fig. \ref{fig:shock} \textbf{A,B}). Parameters describing the transition between edge cell and inner cell behavior also play a role in determining the formation of the invasion band. A narrower transitional zone (smaller \textit{l}) corresponded to a steeper transition between edge and inner cell behavior, which favors the formation of the band. Furthermore, a longer behavioral switching time $\tHat$ leads to increased behavioral variance in the middle of the transitional zone, which also favors the formation of the band (Fig. \ref{fig:shock} \textbf{C,D}).

Finally, we used our model to investigate how disrupting cell behavior leads to changes at the colony level. We consider three types of cell behavior disruption. First, the effect of depolarizing the movement of edge cells, which we find should lead to substantially decreased colony expansion rates and narrower feeding fronts (Fig. \ref{fig:theo}-\textbf{A,B}). Second, we consider the effect of changing doubling time, both by itself but also in conjunction with depolarizing edge cell movement. Here we find that, by itself, lengthening of doubling time should not substantially decrease colony expansion rates, instead just broadening the feeding front (Fig. \ref{fig:theo}-\textbf{A-C}, Fig. \ref{fig:intervention}). However, if lengthening of doubling time is coupled to edge cell movement depolarization, an appreciable decrease in colony expansion rate should occur. Third, we consider the effect of changing cell speed (Fig. \ref{fig:theo}-\textbf{A-C}, Fig. \ref{fig:intervention}), and find that, regardless of edge cell polarization, colony expansion rates should be directly proportional to cell speed (Fig. \ref{fig:theo}-\textbf{C}). Altogether these theoretical results highlight the contribution of edge-cell polarization to the attained colony expansion rate, as well as reveal that such movement polarization might buffer the colony expansion rate from increased cell mortality and disrupted cell division.

\subsection*{Disrupting amoeba cell behavior}

\begin{figure}[ht]
    \centering
\makebox[\textwidth][c]{    \includegraphics[width=6.5in]{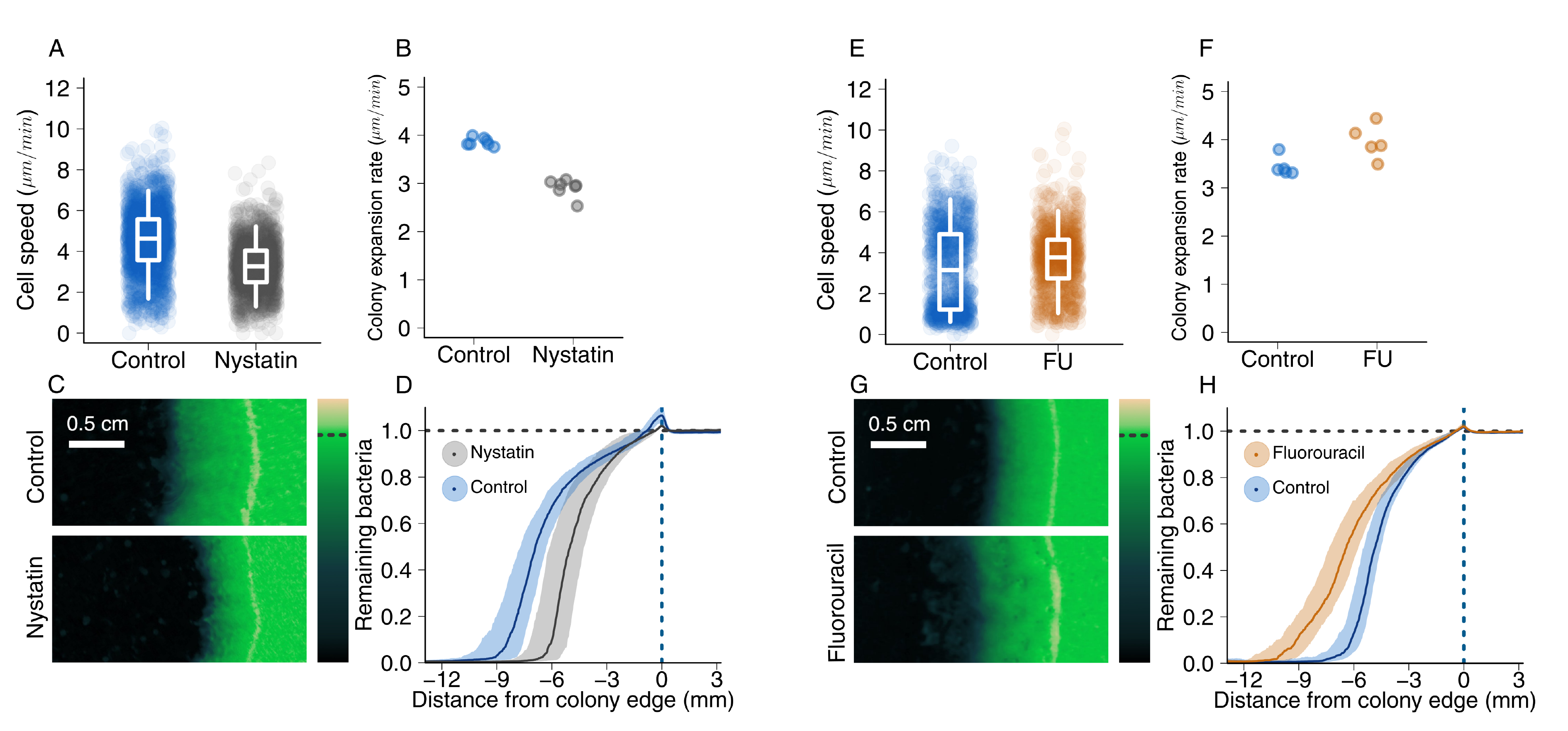}}
\autocaption{Toxin-disrupted \dicty  behavior:}{\textbf{A-D}, Effects of nystatin treatment. \textbf{E-H}, Effects of fluorouracil treatment. \textbf{A,E}, Changes to the median speed of single amoeba trajectories for controls and toxin treatments. Whiskers of the box plots denote the $5^{th}$ and $95^{th}$ percentiles.  \textbf{B,F}, Show the resulting differences to colony expansion rates.  \textbf{C,G}, Representative images of colonies growing under toxin treatments and controls.  \textbf{D,H}, Aggregate profiles of remaining bacteria as a function of distance from colony boundary, both under toxin treatments and controls.}
\label{fig:drugs}
\end{figure}

To probe our theoretical predictions of the colony-level consequences of amoeba cell behavior disruption, we employed toxins that target distinct aspects of amoeba behavior. First, we disrupt cell speed by using nystatin, a toxin naturally produced by the soil-dwelling bacteria \textit{Streptomyces noursei}. Nystatin binds to sterols present in the cell membrane, a common eukaryote-specific target for bacterial toxins, leading to altered cell membrane properties such as increased permeability to ions and increased rigidity\autocite{silva2006competitive}. Although nystatin causes no detectable mortality to amoeba cells in our experimental conditions (Fig. \ref{fig:viability}), the speed of individual edge cells is on average $25\pm 1\% $ slower in the presence of nystatin compared to a control (Fig. \ref{fig:drugs}-\textbf{A}). Consistent with the theoretical predictions, the diminished cell speed in the experiments containing nystatin results in a $25\pm 2\% $ slower expansion rate (Fig. \ref{fig:drugs}-\textbf{B}) as well as in a narrowing of the feeding front (Fig. \ref{fig:drugs}-\textbf{C,D}). Second, we sought to independently disrupt \dicty doubling time by slowing cell division. To this end, we infused our bacterial lawns with fluorouracil, a synthetic uracil analog that acts by inhibiting thymine synthesis to slow down DNA replication and cell division\autocite{loomis1971sensitivity} but leaves the average edge cell speed unaffected (Fig. \ref{fig:drugs}-\textbf{E}). Consequently, consistent with our predictions, we observe no reduction in the expansion rates of fluorouracil treated \dicty colonies (Fig. \ref{fig:drugs}-\textbf{F}), in conjunction with a broadening of the feeding front (Fig. \ref{fig:drugs}-\textbf{G,H}) due to the fluorouracil-induced lengthening of cell doubling times. Overall, our toxin experiments corroborate the distinct theoretically-predicted signatures of disrupting cell movement and doubling time.

\begin{figure}[t!]
    \centering
\makebox[\textwidth][c]{    \includegraphics[width=6.0in]{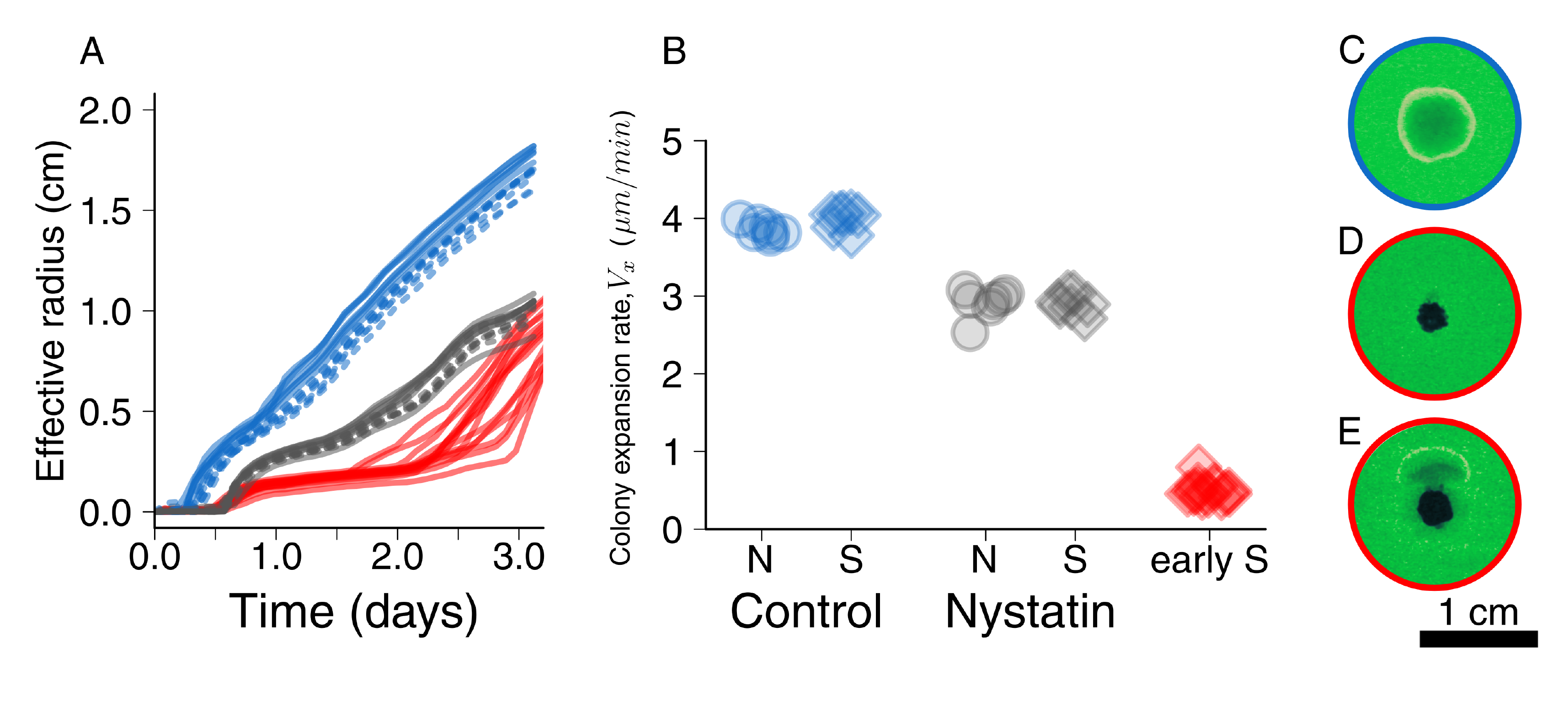}}
\autocaption{Properties of nystatin-tolerant colonies:}{\textbf{A}, Time courses of the effective radius of the bacterial exhaustion zone for \dicty colonies started from cells that were plated immediately after being selected for nystatin-survival (early S cells, red), nystatin-naive (N) cells grown without nystatin (solid blue lines), N cells grown with nystatin (solid gray lines), and cells selected for nystatin-survival that were allowed to grow in nystatin-infused Petri dishes for three days (S cells) grown with (dashed gray lines) or without nystatin (dashed blue lines). \textbf{B}, Colony expansion rates for first two days of growth. \textbf{C-E}, representative images of early colony growth for N cells (\textbf{C}), early S cells (\textbf{D}) and the emergence of naive-like cells in a colony founded by early S cells (\textbf{E}, same colony as \textbf{D}, but a day later).}
\label{fig:evo}
\end{figure}

Toxins might also simultaneously affect both cell movement and doubling times. Although nystatin did not reduce \dicty viability when cells were embedded in the bacterial matrix (Fig. \ref{fig:viability}), high mortality is known to occur for amoeba cells directly exposed to buffer with nystatin. Previous work has shown that it is possible to select for \dicty variants that have lower nystatin-induced mortality, and the resitance mechanism has been linked to membrane sterol compositions with lower nystatin binding\autocite{scandella1980genetic}. Since these resistant variants bind less nystatin, we ask whether the motility effects of nystatin are also  diminished in these resistant cells, which should attenuate the reduction in colony expansion rates caused by nystatin. We repeated our colony expansion experiments with nystatin-resistant amoeba cells, which we selected following previous work\autocite{scandella1980genetic}. Counterintuitively, we find that the nascent colonies seeded from nystatin-survivor cells expand at a much slower rate than those seeded from naive cells, even in the presence of nystatin (Fig. \ref{fig:evo}-\textbf{A,B}). Moreover, our model predicts that such a reduction to the colony expansion rate should be associated with the loss of the distinct invasion band and a strong steepening of the feeding front (Fig. \ref{fig:shock} \textbf{A}, Fig. \ref{fig:theo} \textbf{A}), which indeed we verify experimentally (Fig. \ref{fig:evo}-\textbf{C,D}). 

The slowly expanding, nystatin-survivor phenotype cells are potentially at a selective disadvantage when growing alongside naive cells. Accordingly, in the absence of nystatin, we expect a quick loss of resistance. Indeed, a few days into the expansion of colonies seeded with nystatin-survivor cells, small sectors displaying the naive expansion pattern emerge from points around the colony boundary. These naive-like sectors recover the formation of the broad feeding front with the sharp invasion band, and eventually take over the entire colony boundary (Fig. \ref{fig:evo}-\textbf{D,E}). Subsequently repeating the colony expansion experiments with cells sampled from the naive-like sectors leads to expansion patterns that are indistinguishable from those of colonies seeded with truly naive cells (Fig. \ref{fig:evo}-\textbf{A}), suggesting a phenotypic reversal to the naive state.

Altogether, our results indicate that not only does targeting distinct aspects of cell function lead to distinct responses at the colony level, but that tradeoffs can emerge between different aspects of toxin tolerance. Phenotypes that seemingly provide a viability advantage against a toxin might actually underperform when other aspects of cell function, such as motility, are considered.

\subsection*{Discussion}\label{sec:mistimed}

 When foraging within the bacterial matrix---a three dimensional environment with many structural components---the soil-dwelling amoeba \dicty adopts a spatially-structured foraging strategy with faster, polarized edge cells and slower, unpolarized inner cells. This movement strategy gives rise to fast-expanding colonies that develop a sharp ring of explorer cells. Several distantly related soil-dwelling amoeba species all adopt the same general structured movement mode, leading to colonies with ring-like expansion. Curiously, similar ring-like expansion has been identified in phylogenetically and structurally distinct systems, such as bacteria growing in culture media\autocite{cremer2019chemotaxis}. The pervasiveness of this ring expansion pattern --- which occurs not only across distinct genera of amoebae but also across different domains of life (Eukaryota and Bacteria), regardless of whether cells move using pseudopods or beating flagella, over substrates as distinct as culture media or a bacterial matrix --- might be a product of both the ubiquitous tradeoff between exploiting local resources and exploring the environment\autocite{mehlhorn2015unpacking} and the energetic efficiency of the underlying spatially-differentiated movement modes. 
 
High motility edge cells ensure that the colony will attain a high $\Vx$, while allowing the low motility inner cells to conserve energy without reducing the colony $\Vx$. In the case of amoeba cells, the tradeoff between cell motility and energy efficiency is even more acute than what would be expected solely due the direct energetic costs of locomotion. Because the same cytoskeletal machinery is involved both in pseudopod-based locomotion and in endocytosis, faster moving cells have a slower bacteria intake and accrue less biomass\autocite{chubb2000dictyostelium}. An interesting possibility is that the behavior differences that we observed between inner and edge cells might arise simply by changing the regulation of endocytosis and pseudopod formation. RasS is a protein that is known to be involved in endocytosis regulation, and indeed, \dicty RasS knockouts were shown to have higher rate of pseudopod formation relative to endocytosis rate\autocite{chubb2000dictyostelium}. These mutant lineages display cell speed and polarized morphology very similar to the ones we identified in our wild-type isolated edge cells, suggesting that RasS might be involved in the behavioral differentiation between edge and inner cells. Overall, this observed amoeba cell behavioral differentiation may provide the means to attain a high $\Vx$ while limiting the compromise to bacteria consumption.

The emergent colony expansion rate $\Vx$ could play an important ecological role in the competition among amoebae for bacterial prey. Amoebae inhabit a dynamic soil environment, in which foraging can only occur during short metabolically active periods following rain or nutrient pulses\autocite{clarholm1981protozoan,adl2005dynamics}. To persist, amoebae rely on drought resistant spores that survive and disperse to new bacteria patches during quiescent periods. Amoeba populations compete to produce and disperse these spores. $\Vx$ is likely to affect spore production in two ways: modulating resource consumption and interference with competitors. Firstly, a higher $\Vx$ could allow faster-expanding amoebae to consume more bacteria and thus produce more spores (Fig. \ref{fig:ecological}-\textbf{A,B}). Secondly, if amoebae co-occur in the same bacteria patch, the ones with a higher $\Vx$ may deprive the lower $\Vx$ amoebae of these bacterial resources: faster-expanding amoebae physically displace slower-expanding amoebae at the colony boundary \autocite{buttery2012structured}. This prevents the slower amoebae from accessing bacteria and producing spores (Fig. \ref{fig:ecological}-\textbf{A,B}). Fast expansion (high $\Vx$) thus both increases one's own dispersal capacity and may also reduce the dispersal capacity of others. The relative contributions of these two competitive modes to overall competition depend on ecological factors, including how often competing types co-occur, the durations of metabolically active and quiescent periods, and the size of bacteria patches. Nevertheless, a higher $\Vx$ should always produce a dispersal advantage: higher relative bacteria consumption, more spore production, and consequently higher rates of patch colonization (Fig. \ref{fig:ecological}-\textbf{A,B}).

From the perspective of the bacterial prey, the ability to resist predator invasion throughout the metabolically active soil periods might result in more bacterial cells that can persist throughout the quiescent periods (spores or otherwise), which would provide a population advantage when conditions improve at the start of a new growth cycle (Fig. \ref{fig:ecological}-\textbf{C}). Accordingly, bacteria deploy a series of defenses against invading amoebae: structural elements of the bacterial matrix may hamper the advance of invading amoebae\autocite{huws2005protozoan}, changes to bacteria cell shape impede endocytosis\autocite{jurgens2002predation}, and many toxic secondary metabolites are synthesized\autocite{raaijmakers2012diversity,mazzola2009protozoan,klapper2018role,klapper2016bacterial}. These defenses might act by sterilizing the amoebae and preventing the invasion altogether, by reducing $\Vx$ leading to a decrease in bacteria consumption, or both. However, because the benefit of sterilization without reduction of $\Vx$ is substantially conditioned on total predator elimination, such defense strategies are fragile to predator adaptation or evolutionary change. Point in case, toxin tolerance varies greatly amongst different protozoan predators in the same ecosystem\autocite{brock2016sentinel,klapper2018role}, evolution of resistance often occurs in the face of previously sterilizing toxin doses\autocite{scandella1980genetic}, and toxic secondary metabolites are frequently found in the soil in concentrations that are substantially lower than the experimentally characterized minimum inhibitory concentration\autocite{raaijmakers2012diversity,jousset2008secondary}. This suggests that reduction of $\Vx$ is an important component of bacterial defenses, and our results imply that bacterial defenses that reduce amoeba doubling time --- either by increasing amoeba death rates or decreasing amoeba division rates --- lead to a limited reduction of $\Vx$, whereas defenses that target cell motility but that are not necessarily lethal to the predators --- as we have shown to potentially be the case for nystatin --- achieve a more substantial reduction of $\Vx$. 

If, from an ecological perspective, foraging strategies play an important role in microbial predator-prey dynamics, from an evolutionary perspective these same foraging strategies might have been pivotal to major evolutionary transitions such as the emergence of multicellularity. In particular, dictyostelids have a complex life cycle in which single cells forage until food exhaustion at which point cells aggregate and develop into multicellular fruiting bodies. When first deciphering the mechanisms that allow for cellular aggregation in \textit{D.discoideum}, John Bonner proposed that the cellular machinery responsible for tracking food sources might have been repurposed into the multicellularity pathway\autocite{bonner1969acrasin,bonner2015cellular}. Although it was later established that dictyostelid signalling pathways were not directly derived from prey chemosensing pathways\autocite{shimomura1982chemical}, there is still merit to the general idea that foraging strategies might have provided preadaptations for the emergence of multicellularity. Here we have shown that two important features of multicellular development, namely the possibility of cell behavior differentiation and the occurrence of coordinated cell movement, are present in the unicellular foraging phase of the dictyostelid life cycle. Moreover, these features are also present in feeding fronts of \textit{Acanthamoebae}, a group that diverged from dictyostelids much earlier than the evolution of the multicellular life cycle. This suggests that some pieces necessary for the coordination of multicellular development are ancestral features of amoebozoans, and helps shed light on the recurrent and independent appearances of multicellular life forms across Amorphea, including metazoans, fungi, and slime molds.

\section*{Acknowledgements}
We thank J. Bos for helpful comments and suggestions at various stages of the work. FWR and CET acknowledge support from NSF RoL: FELS: EAGER-1838331. This work was supported in part by the U.S. National Science Foundation, through the Center for the Physics of Biological Function (PHY-1734030), and by National Institutes of Health Grant R01GM097275.

\printbibliography
\appendix

\setcounter{figure}{0}
\renewcommand{\thefigure}{S\arabic{figure}}
\setcounter{page}{1}

\title{\textbf{\SupplementNameFull{} for}\\ \papertitle}

\maketitle

\tableofcontents

\clearpage

\section{Supplemental Figures}

\begin{figure}[ht!]
    \centering
    \makebox[\textwidth][c]{
    \includegraphics[width=6.5in]{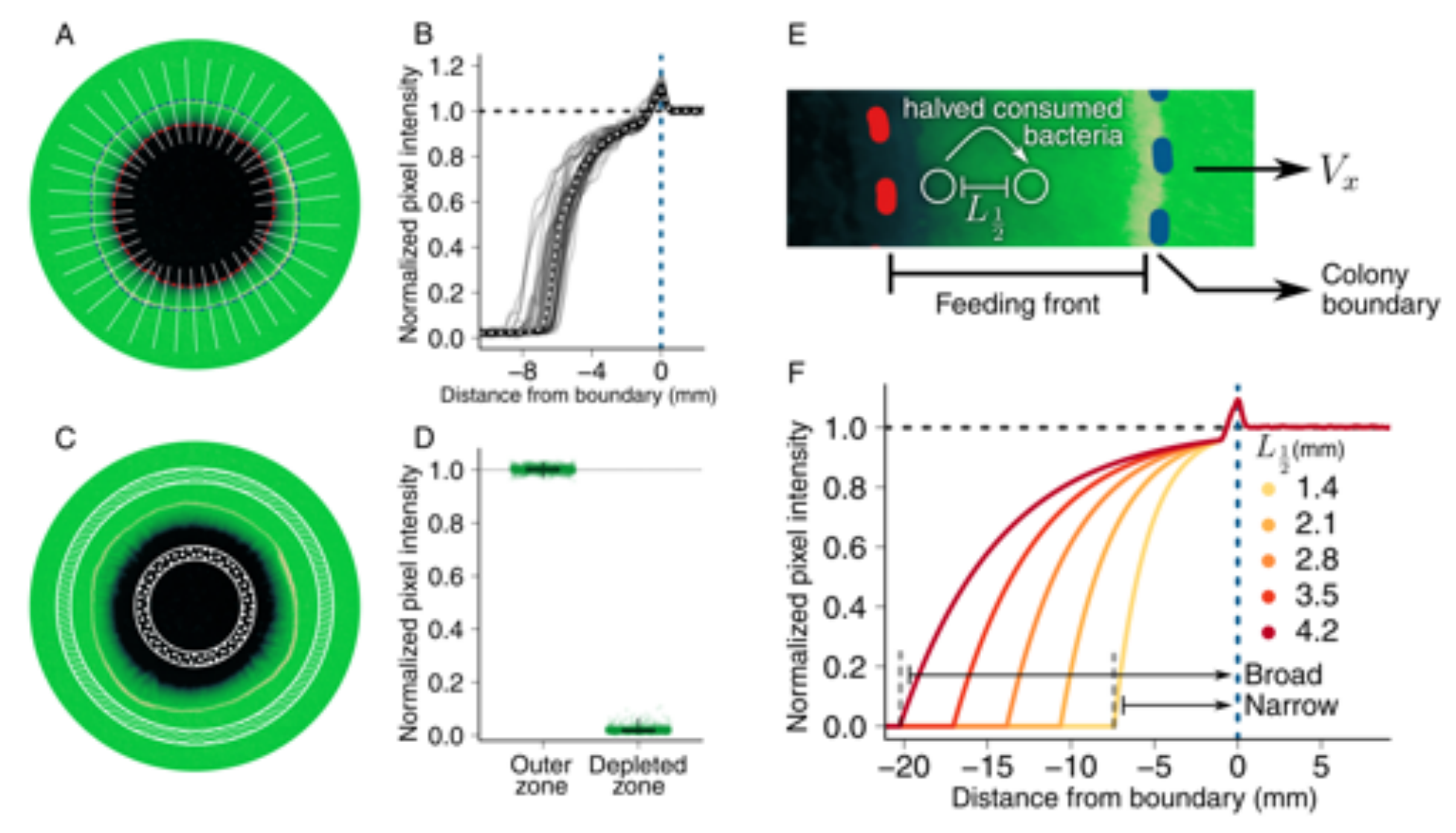}}
    \autocaption{Measuring the amoeba feeding front:}{\textbf{A,} white lines show location of the profiles that were aggregated to construct the average bacteria consumption profiles. \textbf{B} Aggregate bacteria consumption profiles for each of 16 independent Petri dishes. \textbf{C} Outer zone beyond the \textit{D. discoideum} colony boundary and depleted zone in which bacteria have been completely consumed. \textbf{D} Normalized light intensity for pixels in the outer zone shows that there is no artifact of irregular illumination or photobleaching. The low light intensity across the depleted zone shows that any \textit{D. discoideum} autofluorescence is negligible, and that dugestion of bacteria effectively eliminates EGFP fluorescence. \textbf{E} Closeup showing the colony boundary, the feeding front, and the macroscopic quantities $\Vx$ and $\Lhalf$. \textbf{F} Effect of different potential values of $\Lhalf$ on the broadness of the feeding front.}
    \label{fig:measuring}
\end{figure}

\begin{figure}[ht!]
    \centering
    \makebox[\textwidth][c]{
    \includegraphics[width=6.0in]{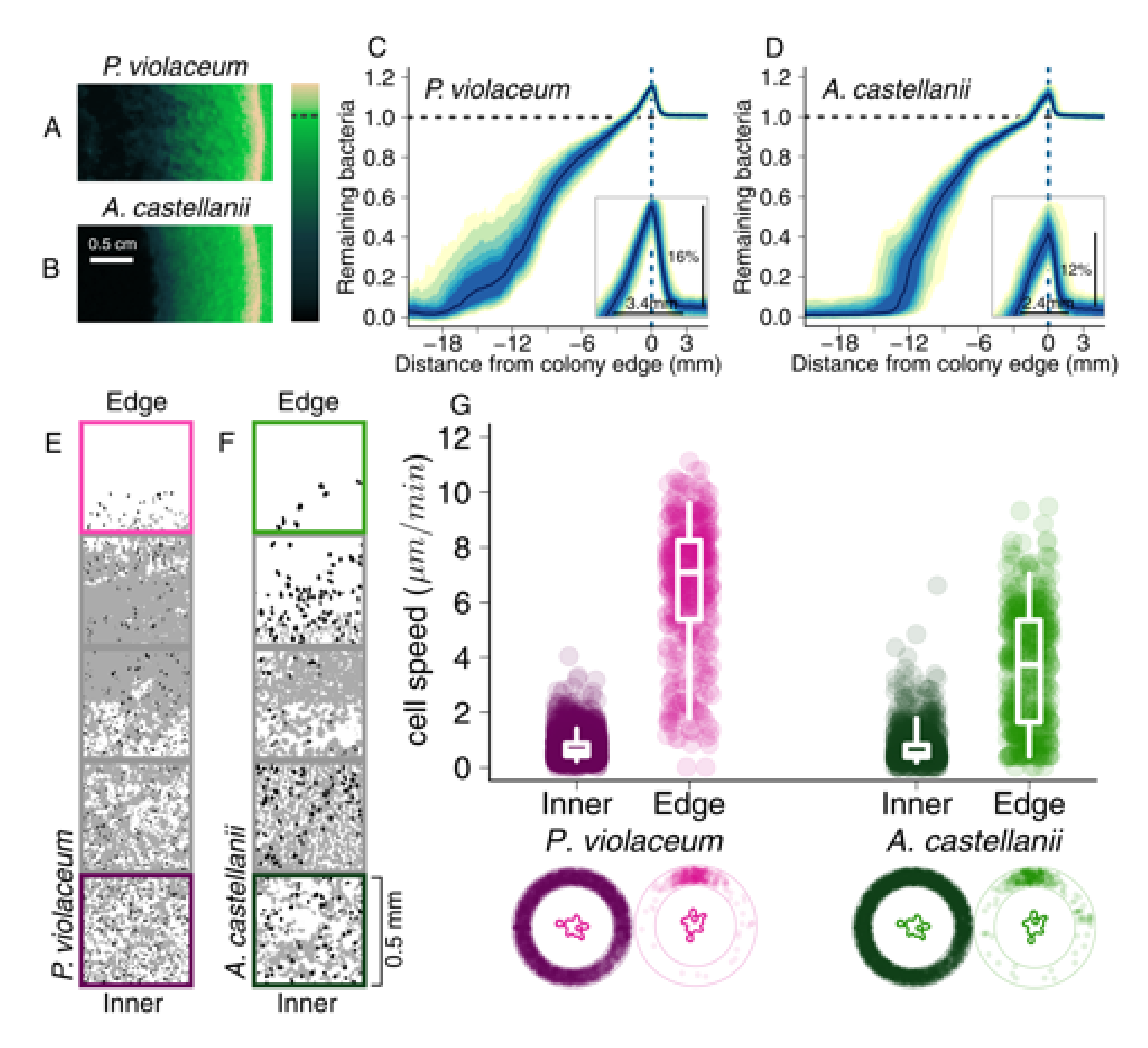}}
    \autocaption{Macroscopic and macroscopic features of \textit{A.castellanii} and \textit{P. violaceum}:}{\textbf{A,B,} Images of feeding fronts for \textit{P. violaceum} (\textbf{A}) and \textit{A. castellanii} (\textbf{B}). False color denotes proportion of remaining bacteria. \textbf{C,D,} Aggregate profiles of remaining bacteria as a function of distance from the colony edge for \textit{P. violaceum} (\textbf{C}) and \textit{A. castellanii} (\textbf{D}), each calculated from 40 profiles taken from each of $15$ dishes. Insets show detail of bacterial accumulation along colony edge. \textbf{E,F,} confocal microscopy views of the colony edge for \textit{P. violaceum} (\textbf{E}) and \textit{A. castellanii} (\textbf{F}). \textbf{G,} Median speed and direction (ring plots) for individual amoeba trajectories, for inner and edge \textit{P. violaceum} and \textit{A. castellanii} cells. Whiskers of the box plots denote the $5^{th}$ and $95^{th}$ percentiles.}
    \label{fig:acan_poly}
\end{figure}

\begin{figure}[ht]
    \centering
    \makebox[\textwidth][c]{
    \includegraphics[width=6.5in]{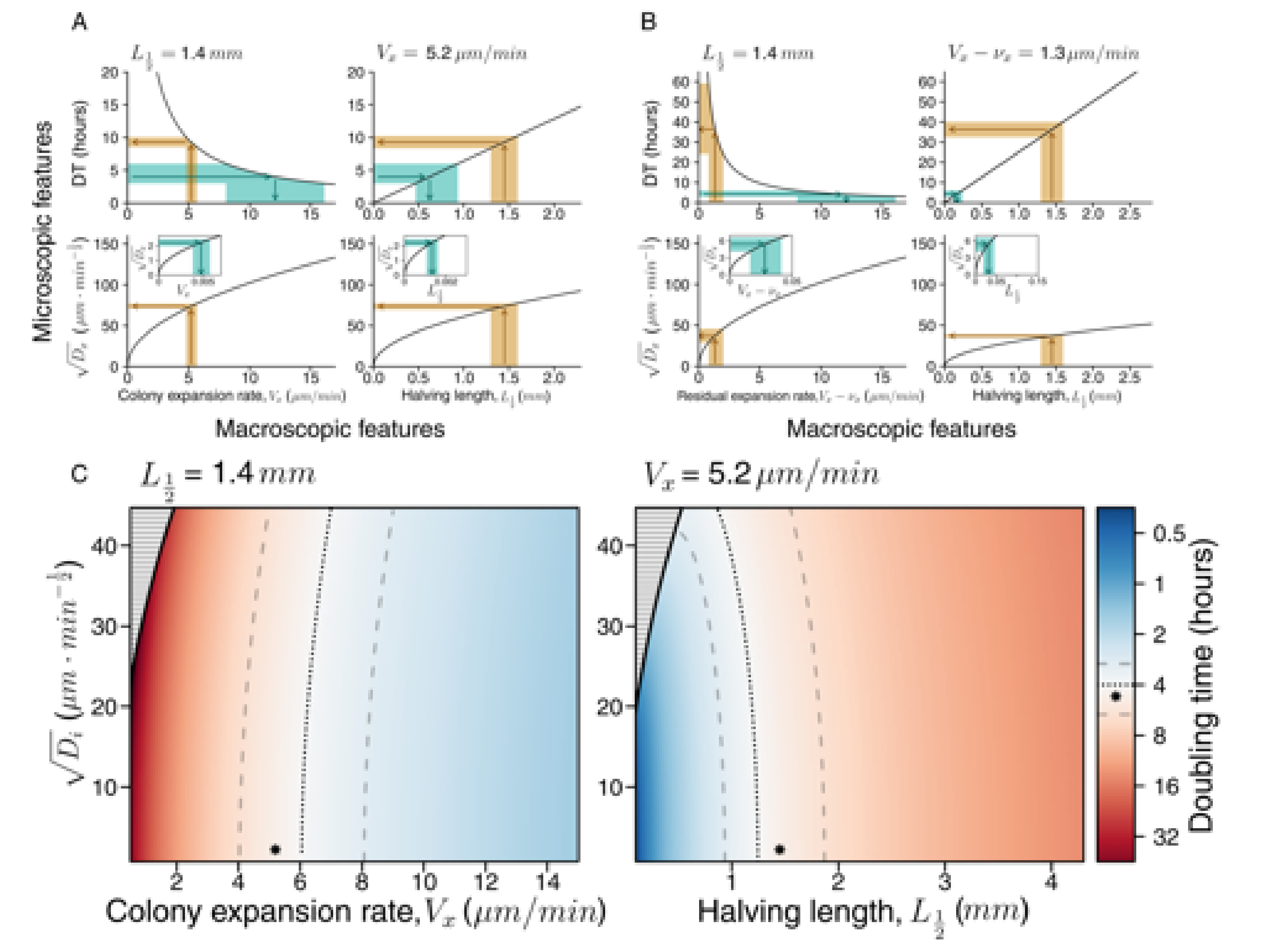}}
    \autocaption{Model agreement with data:}{\textbf{A,B},Black lines show the theoretically predicted relations between macroscopic and microscopic parameters if all cells moved like inner cells (\textbf{A}) or if they all moved like edge cells (\textbf{B}. Copper regions show the experimentally measured macroscopic quantities ($\Vx , \Lhalf$) on the x axis, and the theoretically required matching quantities on the y axis. Teal regions show experimentally measured microscopic quantities on the y axis and the the theoretically required macroscopic quantities on the x axis. The mismatch between copper and teal regions means that under a homogeneous movement model the macroscopic and microscopic data are inconsistent. \textbf{C} Incorporating differential movement rules for inner and edge cells, heatmaps show theoretically predicted doubling times as a function of $\Vx$, $\Lhalf$, and $\Di$. Star shows modal experimental values, and region between dashed lines is the parameter region that leads to doubling times between 2.7 and 5.5 hours (the literature range)}
    \label{fig:agreement}
\end{figure}

\begin{figure}[ht]
    \centering
    \makebox[\textwidth][c]{
    \includegraphics[width=6.0in]{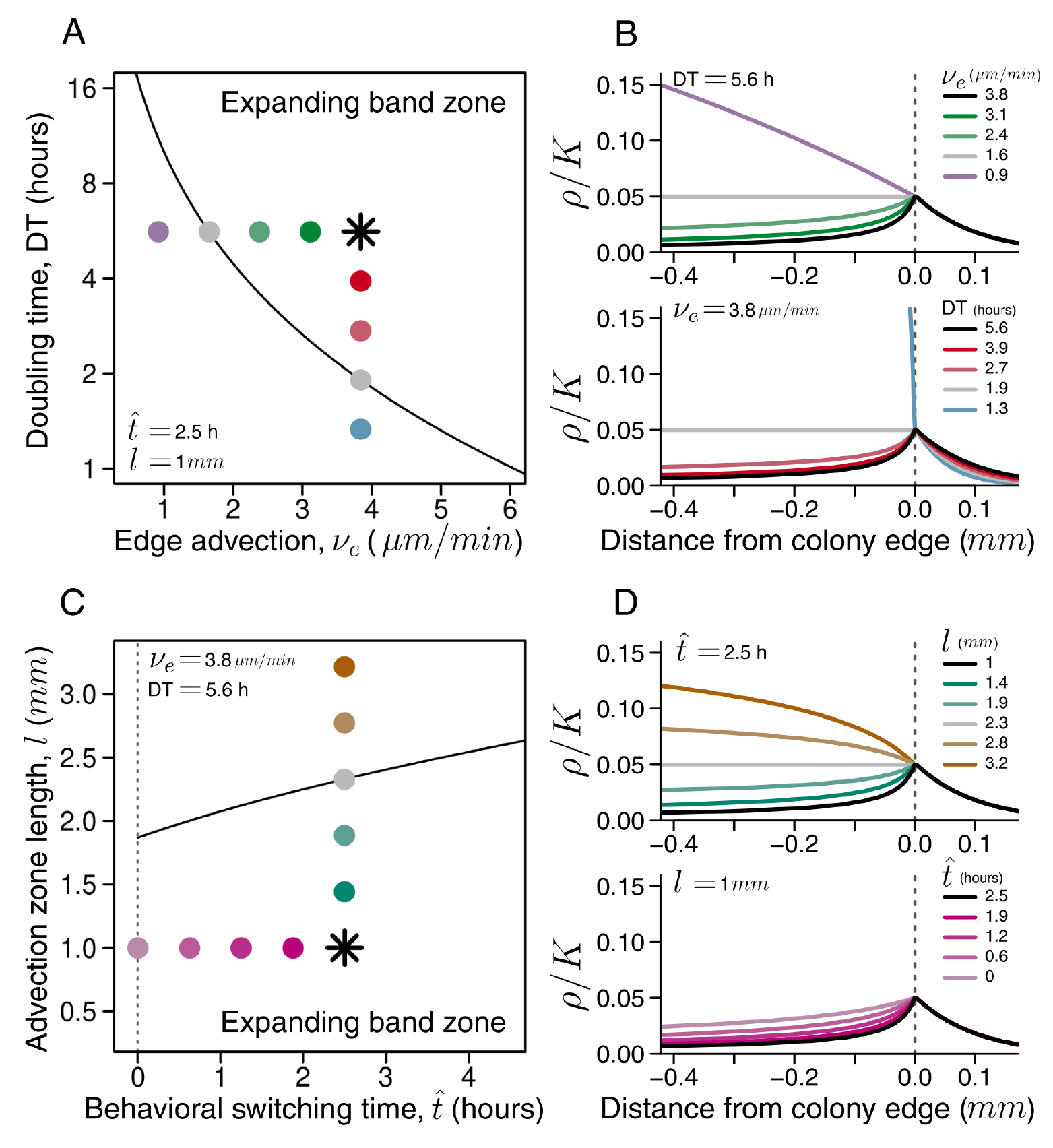}}
    \autocaption{Conditions for the emergence of invasion band:}{\textbf{A,C,} region of the parameter space such that a distinct band forms at the edge of the expansion front for fixed $\hat{t}$ and $l$ and varying $\ve$ and \textsf{DT} (\textbf{A}) and for fixed $\ve$ and $\textsf{DT}$ and varying $\hat{t}$ and $l$ (\textbf{C}). \textbf{B,D,} numerical realizations of the amoeba density around the colony edge for parameter combinations color coded in \textbf{A} and \textbf{C} (\textbf{B} and \textbf{D} respectively).}
    \label{fig:shock}
\end{figure}

\begin{figure}[ht]
    \centering
    \makebox[\textwidth][c]{
    \includegraphics[width=6.5in]{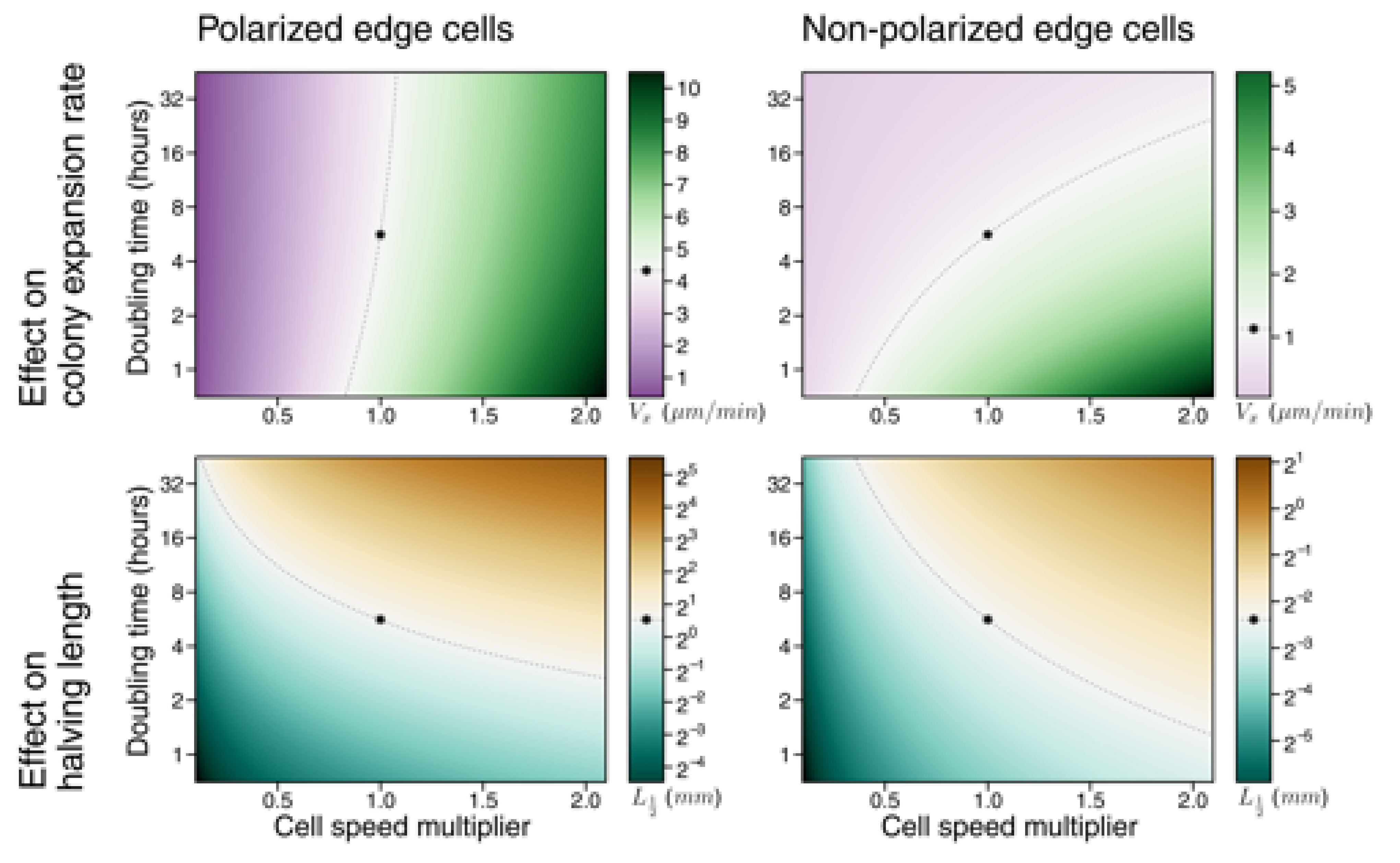}}
    \autocaption{Predicted effects of changing cell speed and doubling time:}{Heatmaps show how jointly changing cell speed and doubling time leads to theoretical changes in halving length and colony expansion rate. Results are shown both incorporating polarized cell movement and in the absence of polarized cell movement.}
    \label{fig:intervention}
\end{figure}

\begin{figure}[ht]
    \centering
    \makebox[\textwidth][c]{
    \includegraphics[width=6.5in]{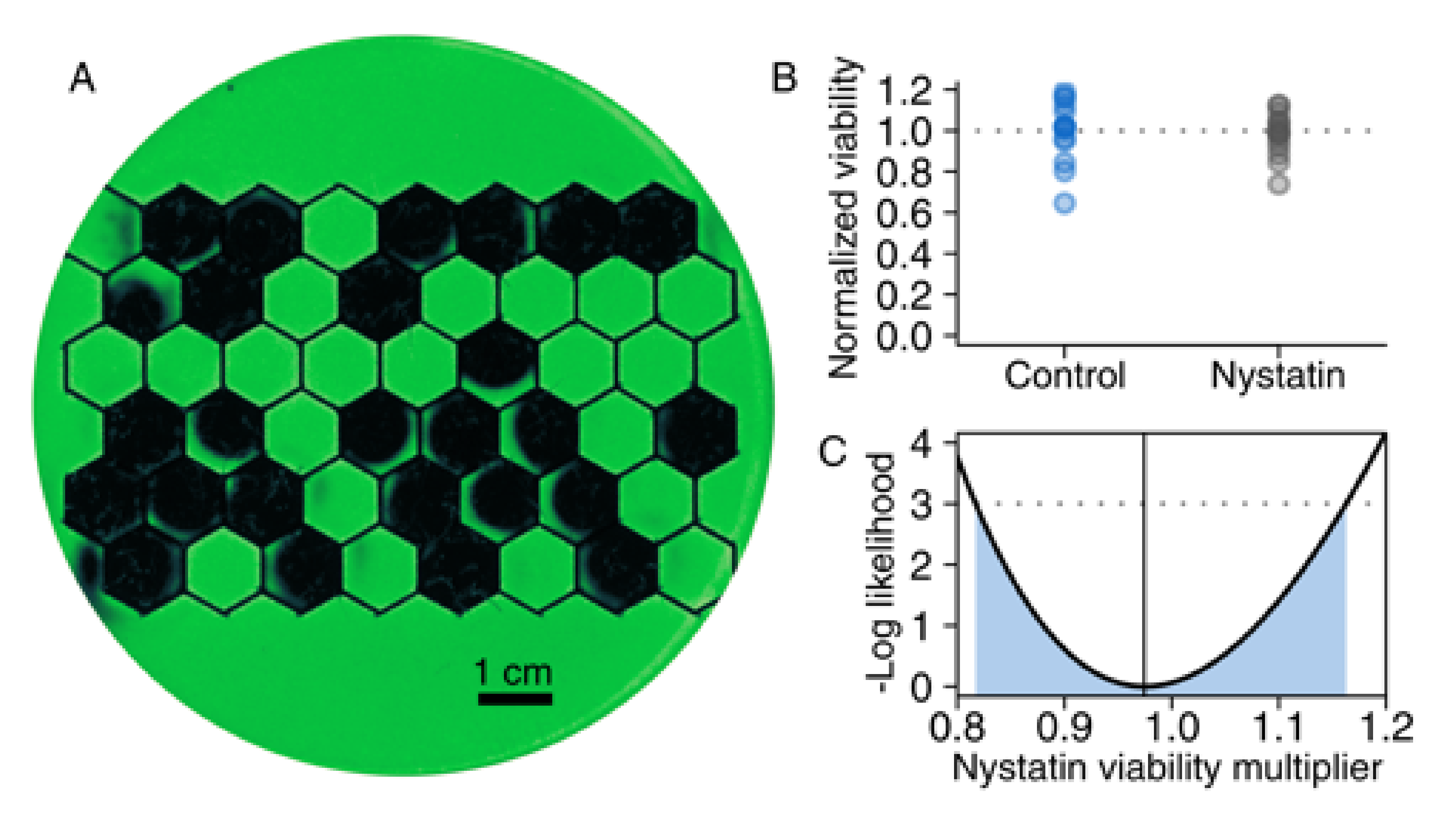}}
    \autocaption{\dicty cell viability under nystatin treatment:}{\textbf{A,} honeycomb lattice cut, nystatin infused Petri dish with \dicty colonies independently growing from single \dicty cells in each hexagonal lattice unit. \textbf{B,} number of lattice units colonized divided by expected number of lattice units that received at least one viable \dicty cell for control Petri dishes and nystatin-infused Petri dishes. Each point is the aggregate count of viable \dicty colonies in $5$ Petri dishes with $48$ lattice units each. \textbf{C} likelihood profile of the effect of the addition of nystatin on \dicty cell viability. Note that the maximum likelihood value is very close to $1$ and that the likelihood ratio interval (shown in blue) contains $1$, denoting no detectable effect of nystatin on \dicty cell viability.}
    \label{fig:viability}
\end{figure}

\begin{figure}[ht!]
    \centering
    \makebox[\textwidth][c]{
    \includegraphics[width=6.5in]{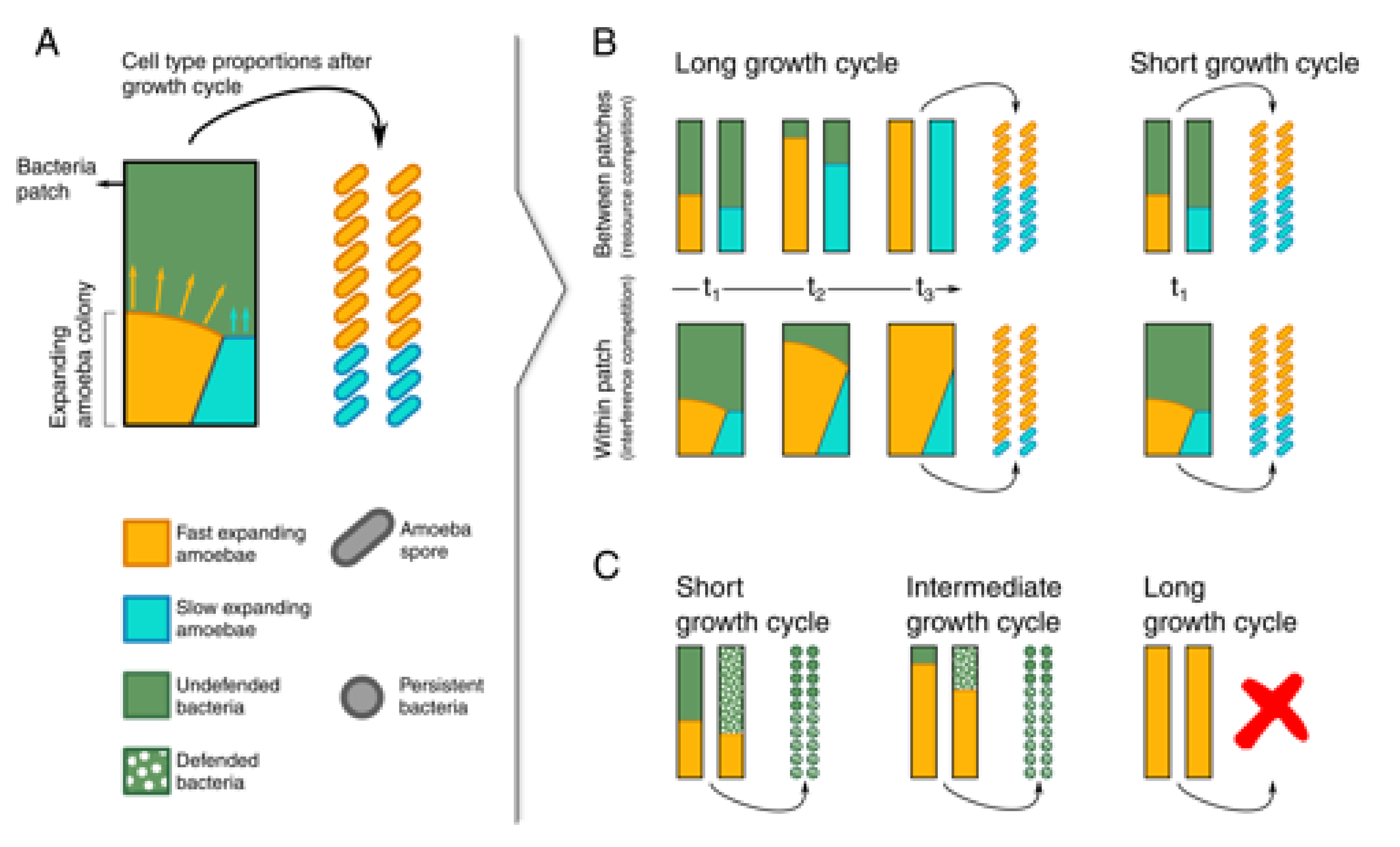}}
    \autocaption{Ecological consequences of varying $\Vx$:}{\textbf{A,} Schematic of a bacteria patch being consumed by high $\Vx$ and low $\Vx$ amoebae. Upper arrow points to relative amount of spores of each amoeba type produced after the growth period, which is proportional to the bacteria area consumed by each amoeba type. \textbf{B,} Schematic for relative amoeba spore production after long growth cycles (shown in three time steps), in which amoebae consume all bacteria, and after short growth cycles, in which some bacteria cells persist after the growth period ends. Upper row shows relative spore production if different amoebae types occur across separate patches (No interference competition). Lower row shows relative spore production of amoebae co-occurring in the same resource patch (interference competition is possible). Note that, if different amoebae co-occur, long growth cycles maximize relative spore differences, whereas, if different amoebae do not co-occur, long growth cycles negate relative spore differences.}
    \label{fig:ecological}
\end{figure}

\clearpage
\section{Methods}

\subsection{Cultures and strains}

Experiments used \textit{D. discoideum} strain NC105.1 sourced from dictybase, \textit{A. castellanii} strain ATCC 30011 from ATCC, and a natural, locally collected \textit{P. violaceum} isolate. All amoebae were grown in \textit{E. coli} lawns in adapted SM $2\%$ agar dishes with only $2.5g/L$ of dextrose. The pH of the agar dishes was measured daily throughout experiments to ensure that it was always between 6 and 7. For imaging, agar had $2\%$ in mass of Mars black pigment (Gamblin) to decrease gel fluorescence, added in prior to autoclaving. The \textit{E. coli} carried the EGFP expressing plasmid pEB1-mEGFP (a gift from Philippe Cluzel via Addgene plasmid 103979 ; http://n2t.net/addgene:103979 ; RRID:Addgene\_103979). Homogeneous \textit{E. coli} lawns were prepared by creating a mist of bacterial culture using an air humidifier and letting the mist settle on the Petri dishes. 

\subsection{Toxins}

Nystatin in DMSO was added to culture media after autoclaving when the temperature fell to 50 degrees celsius to avoid nystatin decomposition. Nystatin concentration was 100 $\mu g/mL$ in treatment samples and DMSO concentration was 2.5 $mL/L$ for both treatment and controls. For the fluorouracil treatments, \textit{E. coli} lawns were grown and then a fluorouracil aqueous mist was let to settle over the agar gel for a final fluorouracil concentration of 100 $\mu g/mL$. It is essential to add fluorouracil after the bacteria lawn has grown, otherwise the drug will inhibit bacterial growth.

\subsection{Macroscopic imaging}

Our fluorescent illumination apparatus consisted of an AAXA P4-X Pico led projector, supplemented with Roscolux CalColor 90 blue cinegel to reduce light bleedthrough. Images were acquired by a DSLR EOS Cannon camera equipped with a Canon RF 24-105mm f/4-7.1 IS STM lens enhanced with a nightsea 600nm emission filter. At each experimental run, up to 30 Petri dishes were sequentially set on an automated machine (built using a Welter’s rectangular lazy Susan as a base) that took photos of each dish at 30 min intervals for a week.

\subsection{Analysis of colony boundaries}

Colony boundaries were determined by leveraging the existence of a highly fluorescent ring around the colony. Microscopic inspection was used to experimentally verify that the ring overlapped with the colony boundary. During image analysis, the most fluorescent closed circumference on the dish was found and its inner area taken to be the amoeba colony. The effective radius of the colony was defined as the radius of a circle with the same area as the colony.

\subsection{Microscopic imaging}

For microscopic imaging, commercial fluorescein-based water tracing dye (EcoClean solutions) was added to media before autoclaving (10 m$L/L$). After Amoeba colonies had grown for about three days, a slice of the gel around the feeding front was removed and placed in a humidified chamber for imaging under a Zeiss LSM microscope. Images were taken for up to 45 minutes after the collection of the slice.

\subsection{Analysis of cell trajectories}

Individual cell trajectories were extracted after image processing using the track mate plugin from Fiji. Confidence intervals for Diffusion and advection coefficients were calculated by bootstrapping: trajectories were resampled with substitution to re-calculate the coefficients. 

\subsection{Cell viability}

To test the effects of nystatin on cell viability we laser cut an hexagonal lattice on the surface of a nutrient agar gel, after bacterial misting, but before lawn growth. After bacterial lawn growth, we added 5 $\mu L$ of \textit{D. Discoideum} cell suspension (at a concentration of approximately one cell per droplet) to each lawn hexagon. Three days later we photographed the Petri dishes and counted the number of viable colonies in treatment versus control. For parameter estimation, colony counts were modelled as a poisson distribution, and cell viability was modelled as a multiplier to the poisson distribution for experiments with nystatin.

\subsection{A model for colony spread}

We model amoebae density $\rho$ as a continuous quantity that changes with amoeba growth and movement. We assume that amoeba behavior is a function of cell position relative to the amoeba colony boundary ($b$). Because of the continuity of our density description, we define $b$ as being the last point in space that has an amoeba density larger than some arbitrary fraction $f$ of a carrying capacity $K$:

\begin{equation}
\begin{aligned}
\label{eqn:boundary_eq}
b(t)=sup\{x\in \mathbb{R}\mid\rho (x,t)> f\cdot K\} 
\end{aligned}
\end{equation}

Putting together growth and movement, we have that

\begin{equation}
\begin{aligned}
\label{eqn:partial_diff}
\frac{\partial }{\partial t}\rho (x,t)=F_x(x,t,\rho)+r\cdot \rho(x,t)\cdot (1-C(K,\rho))
\end{aligned}
\end{equation}

Where $F_x(x,t,\rho)$ is a Fokker-Planck operator that incorporates the randomness of amoeba movement into an diffusion coefficient $D_x (x-b)$ and the biases of amoeba movement into an advection coefficient $\nu_x (x-b)$:

\begin{equation}
\begin{aligned}
\label{eqn:diff}
F_x(x,t,\rho)=-\frac{\partial }{\partial x}\left[ \nu_x (x-b)\cdot \rho(x,t)\right]+\frac{\partial^2 }{\partial x^2}\left[ D_x(x-b)\cdot \rho(x,t)\right]
\end{aligned}
\end{equation}

We assume that amoebae divide at a constant rate $r$ as long as they have bacteria to prey upon:

\begin{equation}
\begin{aligned}
\label{eqn:capacity}
C(K,\rho) = \begin{cases}
0 &\text{if $\rho(x,t) < K$}\\
1 &\text{if $\rho(x,t)) \geq K$}
\end{cases}
\end{aligned}
\end{equation}

We use experimentally measured values of $\nu_x$ and $D_x$ for edge cells ($\nu_e , D_e$) and for inner cells ($\nu_i , D_i$). Following our observation that inner and edge cells retain their differential behaviors even when removed from the feeding front, we allow cells to retain their movement patterns for a certain correlation time $\hat{t}$, after which they reevaluate their behaviors as a function of their position relative to the boundary ($\Bar{x}=x-b$). For simplicity, we say that the probability of a cell behaving as an edge cell decreases linearly from 1 at the colony boundary to 0 at a point inside of the feeding front at a distance $l$ from the boundary ($p_e=1+\Bar{x}/l$). This leads to advection and diffusion coefficients given by

\begin{equation}
\begin{aligned}
\label{eqn:adv_inter}
\nu_x (\Bar{x}) = \begin{cases}
\nu_i &\text{if $\Bar{x} < -l$}\\
\nu_e &\text{if $\Bar{x} > 0$}\\
\nu_i\cdot (1-p_e)+ \nu_e\cdot p_e &\text{if $-l\leq\Bar{x} \leq 0$}
\end{cases}
\end{aligned}
\end{equation}
\begin{equation}
\begin{aligned}
\label{eqn:diff_inter}
D_x (\Bar{x}) = \begin{cases}
D_i &\text{if $\Bar{x} < -l$}\\
D_e &\text{if $\Bar{x} > 0$}\\
D_i\cdot (1-p_e)+ D_e\cdot p_e+\hat{t} \cdot \frac{p_e \cdot (1-p_e )\cdot (v_e-v_i)^2}{2}  &\text{if $-l\leq\Bar{x} \leq 0$}
\end{cases}
\end{aligned}
\end{equation}

Note that the advection coefficient simply linearly interpolates the inner and edge values. However, the diffusion coefficient may have an intermediate maximum. This occurs because within the transitional zone there is a mix of cells behaving either like inner or edge cells, in other words, movement is more unpredictable in the transitional zone. However, if $\hat{t}$ is 0, then cells have no memory and intermediary diffusion is just the linear interpolation.

Importantly, analogously to the derivation of Fisher's equation, we do not derive our equations directly from the integrated stochastic cellular movement and division processes. Instead, we simply add together growth and movement terms that were independently derived. This procedure is known to lead to pathological behaviors for extreme parameter values --- in the case of Fisher's equation, for instance, if growth rates are high enough, one might obtain travelling wave solutions that are faster than the maximum individual speed, which is physically impossible. However, such pathological behaviors only occur for parameter regions far from the experimentally measured quantities, and indeed, individual based simulations indicate that the equations are adequately representing the underlying stochastic process for the parameter range of interest. 

\subsection{Mathematical relations between model quantities}

The standard Fisher's equation --- which differs from our equation in having constant diffusion and advection terms --- has a unique stable travelling wave solution that is asymptotically achieved from initial conditions with compact support (that is, there are infinite possible travelling wave solutions, but only one can be attained if at an initial moment all population density is constrained to a finite region). We do not prove that this property holds for our modified equation, but all of our numerical investigations suggest that this is the case. In this section we assume that this is true and use this property to calculate the $\Vx$ and $\Lhalf$ associated to this asymptotic travelling wave solution.

Since the dynamics around and beyond the colony boundary follow Fisher's equation with constant diffusion and advection coefficients, and since it is the front of the wave that determines wave speed in such cases, following the classical derivation presented by Murray\autocite{murray2007mathematical} the colony expansion speed $V_x$ must be given by equation \ref{eqn:relationsVx}.

To find $\Lhalf$ we separately solve for the back of the wave. Assuming $K$ is large enough, that there is indeed a wave solution with speed $V_x$, and considering that $\nu_i=0$, then, for $\Bar{x}<-l$ equation \ref{eqn:partial_diff} becomes

\begin{equation}
\begin{aligned}
D_i\frac{\partial^2 }{\partial \Bar{x}^2}\rho (\Bar{x}) + V_x\frac{\partial }{\partial \Bar{x}}\rho (\Bar{x})+r\cdot \rho (\Bar{x})= 0
\end{aligned}
\end{equation}

Which has an exponential solution. Since $\Lhalf$ is just $\log2$ divided by the exponent of $\rho$, it can be easily solved leading to equation \ref{eqn:relationsLhalf}.

Finally, to find the conditions for the emergence of a separate invasion band, we note that all we require is that the left hand derivative of $\rho(\Bar{x})$ calculated at 0 be positive. In other words, separation occurs if the amoeba density increases as we approach the colony boundary. A few calculations substituting equations \ref{eqn:diff_inter} and \ref{eqn:adv_inter} into \ref{eqn:partial_diff} lead to the condition that

\begin{equation}
\begin{aligned}
r<\hat{t}\cdot\frac{\nu_e}{l}+\frac{\nu_e}{l}
\end{aligned}
\end{equation}

\end{document}